# Tuning the Hysteresis of a Metal-Insulator Transition via Lattice Compatibility

Y. G. Liang[1], S. Lee[1], H. S. Yu[1], X. H. Zhang[1*], H. R. Zhang[2], L. A. Bendersky[2], Y. J. Liang[3], P. Y. Zavalij[4], X. Chen[5], R. D. James[6], and I. Takeuchi[1*]

[1] Department of Materials Science and Engineering, University of Maryland, College Park, MD 20742

[2] Material Science and Engineering Division, Materials Measurement Laboratory, National Institute of Standards and Technology, Gaithersburg, MD 20899

[3] Chemical and Biomolecular Engineering, University of Maryland, College Park, MD 20742

[4] Department of Chemistry and Biochemistry, University of Maryland, College Park, MD 20742

[5] Department of Mechanical and Aerospace Engineering, Hong Kong University of Science and Technology, Clear Water Bay, Hong Kong

[6] Department of Aerospace Engineering and Mechanics, University of Minnesota, Minneapolis, MN 55455

[*] email: xhzhang@umd.edu; takeuchi@umd.edu

**Abstract**

Structural phase transitions serve as the basis for many functional applications including shape memory alloys (SMAs), switches based on metal-insulator transitions (MITs), etc. In such materials, lattice incompatibility between phases often results in a thermal hysteresis, which is intimately tied to degradation of reversibility of the transformation. The non-linear theory of martensite suggests that the hysteresis of a martensitic phase transformation is solely determined by the lattice constants, and the conditions proposed for geometrical compatibility have been successfully applied to minimizing the hysteresis in SMAs. In this work, we apply the non-linear theory to a strongly correlated oxide system ($V_{1-x}W_xO_2$), and show that the hysteresis of the MIT in the system can be directly tuned by adjusting the lattice constants of the phases. The results underscore the profound influence structural compatibility has on intrinsic electronic properties, and indicate that the theory provides a universal guidance for optimizing phase transforming materials.

## Introduction

The hallmark of first-order structural transformations in solid materials are dramatic changes in physical properties with significant technological implications including caloric effects (1), metal-insulator transitions (2), and enhanced dielectric/piezoelectric susceptibility (3). For metallic alloys, lattice compatibility of the parent phase and the product phase at transformation has proven to be a key factor governing the reversibility of the transition as manifested in the hysteresis of the structural transition (4). Minimization of the hysteresis through tuning of lattice constants, as encoded in the geometrically non-linear theory of martensite (5), has led to development of shape memory alloys with exceptional functional fatigue properties (5-7). In particular, by adjusting the middle eigenvalue $\lambda_2$ of the transformation stretch tensor to 1, a recipe prescribed by the non-linear theory, a precipitous drop in thermal hysteresis was observed. When more stringent conditions (the cofactor conditions) are satisfied (7), a shape memory alloy was found to show unusual domain patterns encompassing multiple length scales and reflecting the ultra-compatibility of the martensite and austenite (5).

Given the ubiquitous nature of first-order transformations, it is of interest to explore the applicability of the non-linear theory of martensite to functional oxide materials: can the brittle ceramic materials also be engineered to have highly-reversible transformations through fine-tuning lattice constants? In this report, we demonstrate that the thermal hysteresis of the metal-insulator transition in W substituted $VO_2$ can indeed be controlled by tuning the lattice constants of the high-temperature tetragonal phase and the low-temperature monoclinic phase.

As an archetypical $3d^1$ correlated oxide, vanadium dioxide ($VO_2$) shows a metal-insulator transition (MIT) at the transition temperature ($T_C$) of $\approx 340$ K (8). The change in the resistivity is generally accompanied by a first-order structural phase transformation between a low-temperature monoclinic phase (M1 phase) and a high-temperature rutile-type tetragonal phase (T phase) (9-12). The relation between the structural phase transformation and the MIT in $VO_2$ has been extensively studied. In particular, many experiments have indicated that the MIT in $VO_2$ is induced by an electron-lattice interaction (i.e., a Peierls transition) through the structural phase transformation (13-15). However, there have been increasing experimental evidence suggesting that the resistance switch and the structural phase transformation in $VO_2$ can be decoupled, and thus the MIT is primarily driven by an electron-electron interaction (i.e., a Mott transition) (16-18). Moreover, in addition to the changes in the structural and the electronic properties, $VO_2$ also shows dramatic changes in many other properties, e.g., the optical transmittance (19), making the material attractive for a number of practical applications, including smart-window coatings, ultrafast sensors, and switching devices (9-12, 19-21). The thermal hysteresis width of pure $VO_2$ is relatively large (> 10 K for a polycrystalline film) (22), which is detrimental to applications requiring agile reversible processes and a large number of reversible cycles.

From the viewpoint of tuning composition to satisfy strong conditions of compatibility between phases, $VO_2$ is an extremely unusual material (7). To explain this assertion, we first note that in general there are two levels of conditions of compatibility known: 1) $\lambda_2 = 1$ and 2) the cofactor conditions. The first of these conditions ($\lambda_2 = 1$) is necessary and sufficient that there is a perfect unstressed interface between phases. The second (cofactor conditions) includes $\lambda_2 = 1$, together with a second condition. The cofactor conditions not only imply perfect unstressed interfaces between T and any single variant of M1, but also imply a huge number of low energy interfaces with any pair of M1 variants, at any volume fraction (See FIG. S1). The two known alloys (5, 23) to accurately satisfy the cofactor conditions have exceptional reversibility of the transformation, including in one case perfect reversibility after 10 million cycles of full stress-induced transformation, under tension, at peak stresses each cycle of 400 MPa.

The "crystallographic accident" of the particular T to M1 transformation in $VO_2$ is that, if $\lambda_2 = 1$ is satisfied, then the cofactor conditions are automatically satisfied (7). Thus, satisfying $\lambda_2 = 1$ in this oxide becomes especially important. In particular, the cofactor conditions are satisfied for the compound twins in this material, of which there are many examples, depending on the choices of variants of the M1 phase. Since we satisfy $\lambda_2 = 1$ in this paper to high accuracy, we here add a third member to list of "cofactor materials".

It is known that the transition properties of $VO_2$ can be effectively tuned through a variety of means, such as chemical doping, electrical field, optical irradiation, external stress, etc. (9-12, 20). Among these approaches, chemical substitution has been extensively investigated (24-28). In particular, tungsten (W) has been reported to reduce the transition temperature by approximately 21 K to 28 K for each atomic percentage of W ions in $VO_2$ (27). However, tunability of the thermal hysteresis width upon increasing the W concentration and its mechanism had not been investigated (29). Through a systematic study using thin-film composition spreads, we show that at $\approx 2.5$ at.% substitution, the middle eigenvalue $\lambda_2$ becomes 1, and the lattice parameters also satisfy the cofactor conditions, resulting in reduced thermal hysteresis width of the metal-insulator transition. Our work underscores the inescapable consequence of lattice compatibility, and signals a unique pathway to control functionalities in a variety of materials including strongly correlated electron systems.

## Results

In this study, a combinatorial film fabrication strategy (30) was adopted to ensure identical deposition parameters for samples with different W concentrations on a given chip. As illustrated in FIG. 1a, continuous composition-spread films of $V_{1-x}W_xO_2$ ($0 \leq x < 4.0\%$) were fabricated by alternatively ablating a $V_2O_5$ target (A) and a $V_{1.92}W_{0.08}O_5$ target (B) using pulsed laser beams; during the ablation of the two targets, an automated moving mask was used to generate a composition gradient across a $c$-$Al_2O_3$ (0001) substrate (henceforth denoted as $V_{1-x}W_xO_2/c$-$Al_2O_3$) or across a Si substrate (henceforth denoted as $V_{1-x}W_xO_2$/Si). The highest W concentration ($x$) in these films was controlled to be less than 4.0% to prevent potential phase separation, which was reported for higher W concentrations (31). Moreover, as illustrated in FIG. 1b, each composition-spread film was patterned into multiple parallel strips (each with a length of 5 mm and a width of 0.2 mm) along the composition gradient direction for further measurements.

The XRD spectra of the sample strips across the composition-spread films were measured at different temperatures. It is important to note that on both $c$-$Al_2O_3$ and Si substrates, only peaks associated with the monoclinic phase and/or the tetragonal phase of $V_{1-x}W_xO_2$ could be identified in the entire scan range ($10°$ $\leq 2\theta \leq 90°$), indicating the absence of other phases of vanadium oxides and tungsten oxides.

Furthermore, the XRD results obtained from $V_{1-x}W_xO_2$/Si samples had shown a number of diffraction peaks revealing the polycrystalline nature of these films. Among the peaks observed in the diffraction scans, the one appearing at around $28.0°$ shows strong intensity, thus allowing the study of the phase evolution. FIG. 2 shows the XRD spectra of a $V_{1-x}W_xO_2$/Si sample in a range from $26°$ to $31°$ at different temperatures.

As shown in FIG. 2a, the spectrum obtained at 300 K from each of the four sample strips with $x < 1.5\%$ clearly shows a peak around $28.0°$, which is consistent with the (011) of the M1 phase; the scans obtained from the strips with $x > 1.5\%$ shows a peak around $27.75°$, which is the (110) of the T phase. Moreover, the spectrum obtained from the sample strip with $x = 1.5\%$ shows a relatively broad peak, manifesting the coexistence of the M1 phase (the peak at around $28.00°$) and the T phase (the peak at around $27.75°$). Therefore, by increasing the W concentration, the insulating M1 phase at 300 K gradually evolves into the metallic T phase. Compared to the slight shift of the peak towards a lower angle within the M1 phase or within the T phase entirely due to smaller V atoms replaced by bigger W atoms (29, 32-33), the shift of the diffraction peak is more prominent when the crystal structure changes from the T phase to the M1 phase at a critical W concentration of about 1.5%.

As shown in FIG. 2b, at 323 K, the position of the peak obtained from each sample strip with $x > 1.5\%$ remains unchanged. The peak for the sample strip with $x = 1.5\%$ becomes sharper and clearly centered at about 27.75°, indicating that the sample strip is fully in the metallic T phase. Compared to that observed at 300 K, the peak for the sample strip with $x = 0.9\%$ at 323 K appears at an angle 0.25° smaller, suggesting that a transformation from the M1 phase to the T phase is completed as the temperature increases. Moreover, a clear double-peak feature observed in the spectrum obtained from each of the three sample strips with $x < 0.9\%$ indicates the coexistence of the two phases at 323 K.

As shown in FIG. 2c, when the temperature is further increased to 358 K, all the sample strips show a sharp peak at around 27.75° in the XRD spectra. Therefore, with $x$ in the range of 0 - 3.4%, the entire composition-spread $V_{1-x}W_xO_2$ film is in the high-temperature metallic T phase. The slight shift of the peak towards a lower angle as $x$ increases is an indication of size effect of the substitution of V with W in the $V_{1-x}W_xO_2$ system (29, 32-33).

In contrast to the multi-peak polycrystalline XRD results obtained from the $V_{1-x}W_xO_2$/Si sample, the XRD results of all the sample strips in a $V_{1-x}W_xO_2$ film fabricated on a $c$-$Al_2O_3$ substrate show only two peaks belonging to the film at about 40.00° and about 86.30° (FIG. 3a). The positions of the two peaks correspond to the (020) and the (040) of a slightly W-doped $VO_2$ film in either the M1 phase ($b_{M1} = 4.52$ Å) (ref. 20) or the T phase ($a_T = 4.55$ Å) (ref. 20). Therefore, the results suggest that the $V_{1-x}W_xO_2$ film was epitaxially grown on the $c$-$Al_2O_3$ substrate following the (020)//(0006) relationship between $V_{1-x}W_xO_2$ and sapphire (34). However, because other diffraction peaks are not visible, the XRD results obtained for the epitaxial $V_{1-x}W_xO_2$ film alone are not sufficient to provide evidence for the doping induced phase transformation.

In order to further investigate the doping effect in the composition-spread $V_{1-x}W_xO_2$/$c$-$Al_2O_3$ samples, scanning transmission electron microscopy (STEM) measurements were carried out on cross-sectional samples. A representative TEM image of the $V_{1-x}W_xO_2$/$c$-$Al_2O_3$ heterostructure at $x = 0\%$ is shown in FIG. 3b; the $V_{1-x}W_xO_2$ thin film has a near-uniform thickness of 150 nm and is composed of columnar grains. FIG. 3c shows a selected area electron diffraction pattern (SAEDP) taken from the film with $x = 0\%$. The SAEDP can be indexed with two variants of the M1 phase along [102] and [201] zone-axes: the alignment of the variants is consistent with the pseudo-hexagonal symmetry of the monoclinic $VO_2$. Furthermore, as shown in FIG. 3d, the atomic resolution high-angle annular dark-field (HAADF) STEM image taken from the film at $x = 3.44\%$ indicates the coexistence of the T phase and the M1 phase. Therefore, although it is difficult to resolve the M1 phase and the T phase by XRD on $c$-$Al_2O_3$ (FIG. 3a), the STEM results clearly indicate that increasing the W concentration leads to formation of co-existing M1 and T phases (27).

In lightly-doped $V_{1-x}W_xO_2$ films, the V/W ions in the high-temperature tetragonal T phase are aligned along the $c$-axis of the crystal structure (35). The structural phase transformation to the low-temperature M1 phase and, accordingly, the symmetry breaking results in the observing the V/W ions aligned into zigzagging chains (36). The inset of FIG. 4 illustrates the correspondence between two tetragonal unit cells in the T phase and one monoclinic unit cell in the M1 phase through the structural transformation. It is important to note that, for illustrative purposes, the unit cells shown in the inset of FIG. 4 merely reflect the correspondence of the lattice parameters in the two end phases without taking the lattice stretching effects into account. In fact, the actual lattice constants, which can be determined from the XRD results, are the key to the interfacial stress generated during the phase transformation.

To limit the interfacial stresses which develop during the structural phase transformation, high compatibility between the two end phases is required. According to the non-linear theory (5-7), when the cofactor conditions are satisfied, a highly compatible phase transformation becomes possible, resulting in the minimal thermal hysteresis width (4). Because the transformation stretch matrix is solely determined by the structural relationship between the two end phases, the X-ray measurements and Rietveld refinement (6) were used to establish the crystal symmetries and lattice constants for different W concentrations across the composition-spread films. Table S1 in Supplementary Information shows a list of the lattice parameters obtained at different W concentrations in the two end phases.

A common geometric feature of the stretch tensors in Table S1 is that the middle eigenvalue $\lambda_2$ is associated with the eigenvector aligned with the tetragonal a-axis. In this case, the kinematic compatibility conditions (7) ensure that when the lattice parameters are tuned to satisfy $\lambda_2 = 1$, a laminate of monoclinic compound twins is compatible with the tetragonal phase at arbitrary volume fraction of the twin pair. This is illustrated in FIG. S1, using our measured lattice parameters at $x = 2.4\%$. Note the excellent matching of phases despite quite large distortions of about 6% strain. In particular, in the 1st and 6th schematic pictures of FIG. S1 (i.e., FIG. S1a and S1f), a single monoclinic variant matches the tetragonal lattice at a stress-free interface without any transition layer. Again using our measured lattice parameters of $V_{1-x}W_xO_2$ at $x = 2.4\%$, we blow up and plot the local structure of the interface in the region of the circle. The interface normal in this case is (0., 0.507361, 0.861733)t.

The complete analysis of the measured lattice parameters at different W concentrations provides the middle eigenvalue $\lambda_2$ of the transformation stretch matrix as a function of the W concentration (FIG. 4). We find that the $b_{M1}/a_T$ ratio (also shown in FIG. 4) is highly correlated with the middle eigenvalue $\lambda_2$ for all W concentrations, both approaching 1 simultaneously at a W concentration of about 2.4%. As shown in the inset of FIG. 4, the $b_{M1}$ lattice is not only normal to both $a_{M1}$ and $c_{M1}$ lattices in the M1 phase, but also corresponds to one of the $a_T$ lattices of the tetragonal unit cell of the T phase. The fact that the $b_{M1}/a_T$ ratio is always the middle eigenvalue $\lambda_2$ suggests that the stretch or compression through the structural phase transformation is mainly perpendicular to $b_{M1}$. Furthermore, the results also suggest that any stretch or compression along the direction of the $b_{M1}$ lattice would enhance the deformation in directions perpendicular to the $b_{M1}$ lattice. Therefore, only when the $b_{M1}/a_T$ ratio becomes 1 (i.e., the cofactor conditions are satisfied) at $x = 2.4\%$, the structural deformation in directions perpendicular to the $b_{M1}$ lattice is minimized, and phase transformation with minimal distortion becomes possible.

According to the non-linear theory (5-7), the fulfillment of the cofactor conditions is expected to minimize the interfacial energy involved during the phase transformation, and thus reduce the width of the hysteresis loop upon thermal cycling. Therefore, the thermal hysteresis width is expected to provide another measure for the compatibility between the two end phases. Among various physical properties of $V_{1-x}W_xO_2$ that show a thermal hysteresis loop, the MIT provides a convenient route to quantify the transition temperature $T_C$ and the thermal hysteresis width $\Delta T$ (defined below).

Electrical measurements were performed on both epitaxial and polycrystalline films to investigate the effect of W doping on the characteristics of MIT. FIG. 5a and 5b show the hysteretic temperature-dependent sheet-resistance curves ($R_S$-T) of composition-spread thin films on $c$-$Al_2O_3$ and on Si, respectively. For both composition-spread films, the expected systematic reduction of the transition temperature with increase in the W concentration is seen. Specifically, as shown in FIG. S2, the transition temperature shows a nearly linear decrease with the increase of W concentration for both films. The linear fits shown in FIG. S2 further suggest that corresponding to each at.% increase in the W concentration, the transition temperatures in the epitaxial film and the polycrystalline film are reduced by 25 K and 21 K, respectively, which agree well with the values reported previously (37).

FIG. S3 provides an example of how we extract the $T_C$ and the thermal hysteresis associated with the transition $\Delta T_C$ from a measured $R_S$-T curve. The $R_S$-T curve obtained on a $VO_2$ (i.e., $x = 0$%) strip fabricated on a $c$-$Al_2O_3$ substrate shows a change of nearly 4 orders of magnitude through the MIT. Two transition temperatures ($T_{Cup}$ and $T_{Cdn}$) were obtained by taking the temperatures at the dips in the first order derivatives of the warming curve and the cooling curve, respectively (38). Further, a phase transition temperature $T_C$ at about 342 K, consistent with that obtained from $VO_2$ single crystals (19), was determined as the average of the two transition temperatures, i.e., $T_C = (T_{Cup}-T_{Cdn})/2$. The hysteresis width $\Delta T_C$ is defined as the difference between the two transition temperatures, i.e., $\Delta T_C = T_{Cup}-T_{Cdn}$, which for this strip is 6.86 K.

In FIG. 5c, the measured $\Delta T_C$ was plotted as a function of W concentration for both films. The overall hysteresis width $\Delta T_C$ for the $V_{1-x}W_xO_2$/Si sample is clearly larger than that for the $V_{1-x}W_xO_2$/$c$-Al$_2$O$_3$ sample. The difference in the values of the hysteresis width $\Delta T_C$ may be attributed to the polycrystalline nature of the film grown on the Si substrate (35, 39). Despite the difference in the exact values of the measured $\Delta T_C$, both spread films share a common feature – there is a clear drop in $\Delta T_C$ at the W concentration of $\approx 2.4\%$, indicating the importance of satisfying the cofactor conditions in minimizing the thermal hysteresis width. Furthermore, as shown in FIG. 5d, the measured $\Delta T_C$ for the $V_{1-x}W_xO_2$/Si sample is also plotted as a function of $\lambda_2$. Clearly, as $\lambda_2$ becomes 1 (i.e., the cofactor conditions are satisfied), the thermal hysteresis width $\Delta T_C$ reaches the minimum value, thus confirming that fulfilling the cofactor conditions indeed leads to minimization of the thermal hysteresis width $\Delta T_C$. Therefore, as evident in the composition-spread $V_{1-x}W_xO_2$ films grown on different substrates, the non-linear theory of martensite is applicable not only to metallic systems but also to oxide systems.

## Discussion

In this study, high-quality composition-spread $V_{1-x}W_xO_2$ films were fabricated on $c$-Al$_2$O$_3$ and Si substrates using a high-throughput pulsed-laser deposition technique. X-ray diffraction and electronic transport measurements were performed to systematically investigate the effect of W doping on the structural phase transformation and the metal-insulator transition (MIT). Based on the lattice parameters determined from the X-ray diffraction measurements at different temperatures, we found that the cofactor conditions based on the geometrically non-linear theory of martensite (GNLTM) are satisfied at a W concentration of 2.4%. The measurements of the metal-insulator transition in the spreads indicate that the thermal hysteresis width indeed reaches the minimum value for the samples with W concentration near 2.4%. The success of applying the non-linear theory to identify the conditions for the ultra-compatible MIT in a functional oxide system suggests that the theory can provide a universal guidance for optimization of transforming materials in general. Moreover, the correlation between the MIT and the crystal structure demonstrated in our study also suggests that the structural phase transformation plays a central role in the observed MIT in the $V_{1-x}W_xO_2$ system.

## Materials and Methods

The continuous composition-spread films of $V_{1-x}W_xO_2$ ($0 \leq x < 4.0\%$) used in this study were fabricated in a combinatorial pulsed-laser deposition (PLD) chamber. During the deposition, the energy of the laser pulses was 18 mJ, the frequency of the pulses was 5 Hz, the oxygen pressure was about 3 mTorr, and the substrate temperature was about 500 °C. Moreover, in order to minimize the substrate-induced strain, the composition-spread films used in this study all had a thickness of ≈150 nm. Wavelength dispersive X-ray spectroscopy (WDS) was performed to characterize the W concentration in different positions of the composition-spread $V_{1-x}W_xO_2$ films.

The X-ray diffraction (XRD) measurements were carried out in a Bruker D8 Discover system. The system was equipped with an area detector and a stage that allows the sample's translation in the direction of a composition gradient for automated data collection. With in-situ temperature control of the sample using either a heater or a liquid-nitrogen cold bath attached to the sample stage, θ-2θ X-ray spectra were collected at several temperatures from 255 K to 358 K.

The temperature dependence of the electrical resistance was measured for each sample strip in the $V_{1-x}W_xO_2$ spread films by utilizing a four-probe geometry using a physical property measurement system (PPMS) made by Quantum Design Inc. To improve the measurement accuracy, temperature scans were all carried out with a sweeping rate of 1.0 K/min and a measurement step of 0.2 K.

## Acknowledgements

This work was supported by ONR MURI N000141310635 and ONR MURI N000141712661 at UMD. XC thanks the HK Research Grants Council for financial support under Grants 26200316 and 16207017. RDJ was supported by NSF (DMREF-1629026), ONR (N00014-18-1-2766), MURI (FA9550-18-1-0095) and a Vannevar Bush Fellowship. He also benefited from the support of Medtronic Corp, the Institute on the Environment (RDF fund), and the Norwegian Centennial Chair Program. XC and RDJ thank the Isaac Newton Institute for Mathematical Sciences for support and hospitality during the program "The mathematical design of new materials" (EPSRC EP/R014604/1) when work on this paper was undertaken.

**Figure Legends**

FIG. 1  Fabrication of a continuous composition-spread film of $V_{1-x}W_xO_2$. (a) A schematic diagram of the set-up for film growth; (b) A schematic view of a patterned spread film.

FIG. 2  XRD θ-2θ patterns obtained from different sample strips of a $V_{1-x}W_xO_2$ film grown on a Si substrate at several representative temperatures: (a) 300 K, (b) 323 K, and (c) 358 K.

FIG. 3  (a) XRD θ-2θ patterns obtained from different sample strips of a $V_{1-x}W_xO_2$ ($0 \leq x < 4\%$) film grown on a $c$-$Al_2O_3$ (0001) substrate at room temperature; (b) a cross-sectional STEM image of the $V_{1-x}W_xO_2$/$c$-$Al_2O_3$ film; (c) a typical selected-area electron diffraction pattern (SAEDP) taken from the pure $VO_2$ side of the $V_{1-x}W_xO_2$/$c$-$Al_2O_3$ film; and (d) an atomically-resolved HAADF-STEM image of the $V_{1-x}W_xO_2$/$c$-$Al_2O_3$ film at a position with $x = 3.44\%$.

FIG. 4  W concentration ($x$) dependence of the middle eigenvalue $\lambda_2$ of the stretch matrix (empty circles) and the $b_{M1}/a_T$ ratio (solid dots), respectively; the inset illustrates a schematic transformation from two tetragonal unit cells in the T phase to one monoclinic unit cell in the M1 phase, where $a_T$ and $c_T$ are the lattice parameters of the tetragonal unit cell, while $a_{M1}$, $b_{M1}$, $c_{M1}$, and $\beta$ are the lattice parameters of the monoclinic unit cell.

FIG. 5  The temperature dependence of the resistance ($R_s$) measured from sample strips at different W concentrations in a $V_{1-x}W_xO_2$ film grown on $c$-$Al_2O_3$ (a) and on Si (b). (c) the W content ($x$) dependence of hysteresis width ($\Delta T_C$) for both $V_{1-x}W_xO_2$/$Al_2O_3$ and $V_{1-x}W_xO_2$/Si films; (d) hysteresis width ($\Delta T_C$) as a function of $\lambda_2$ for $V_{1-x}W_xO_2$/Si.

**Figures**

(a)

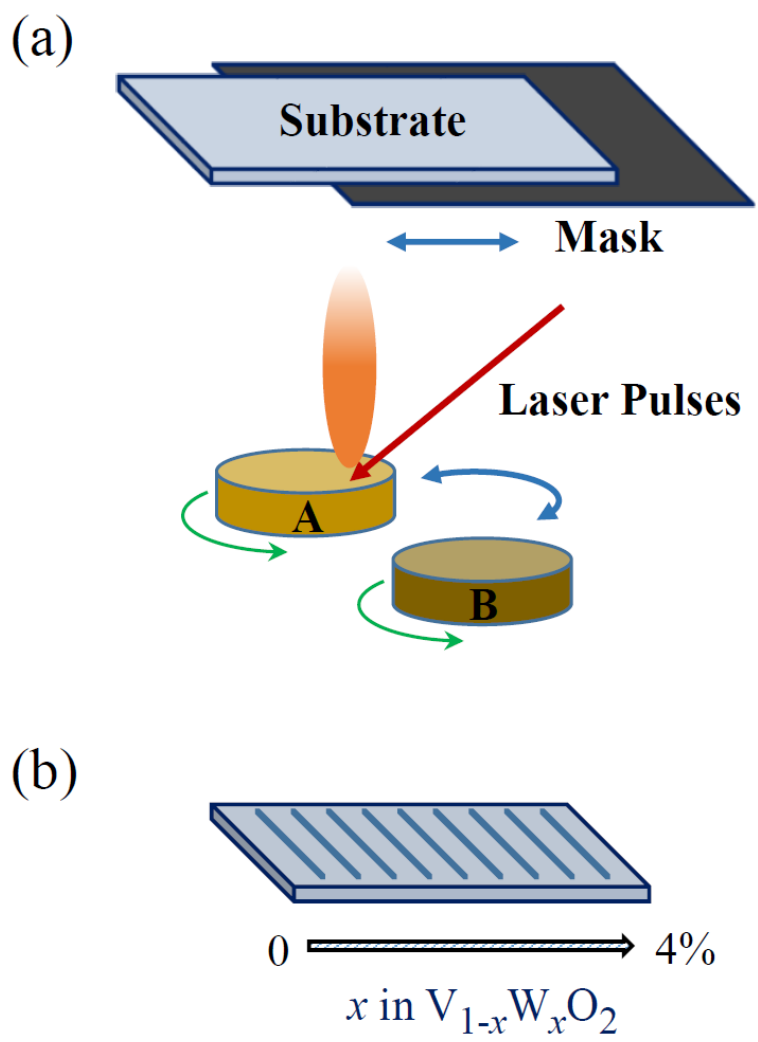

(b)

FIG. 1

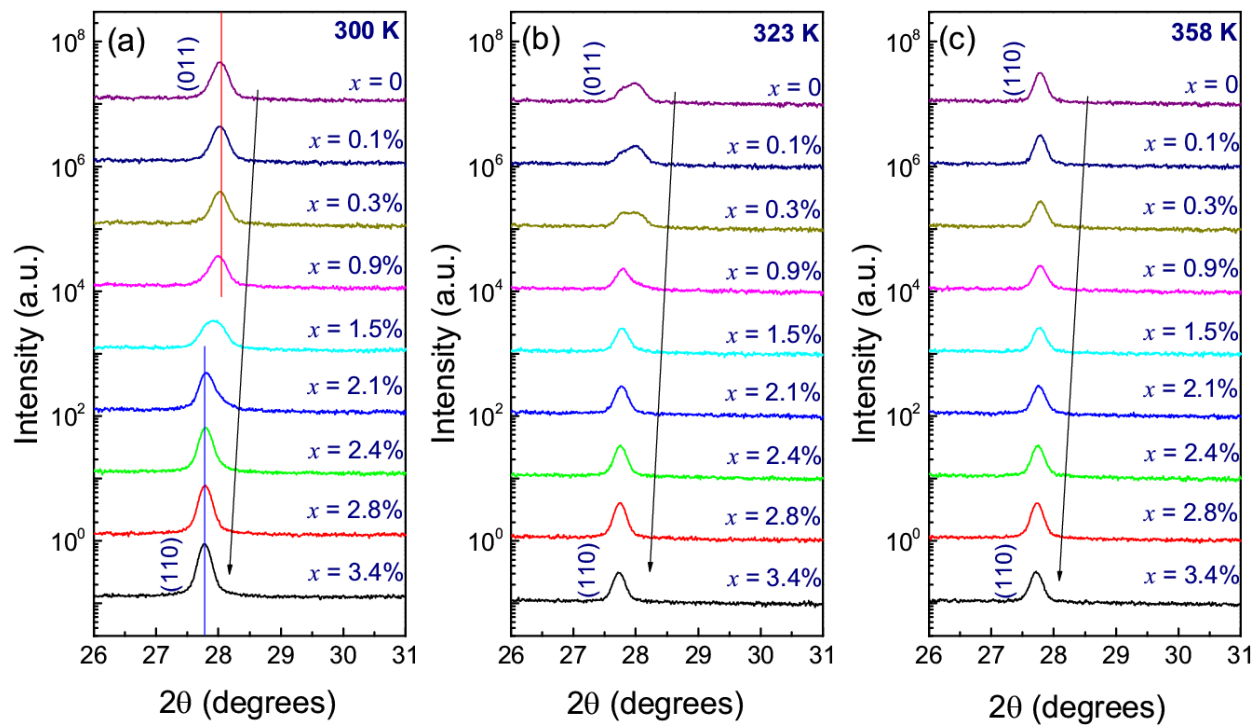

FIG. 2

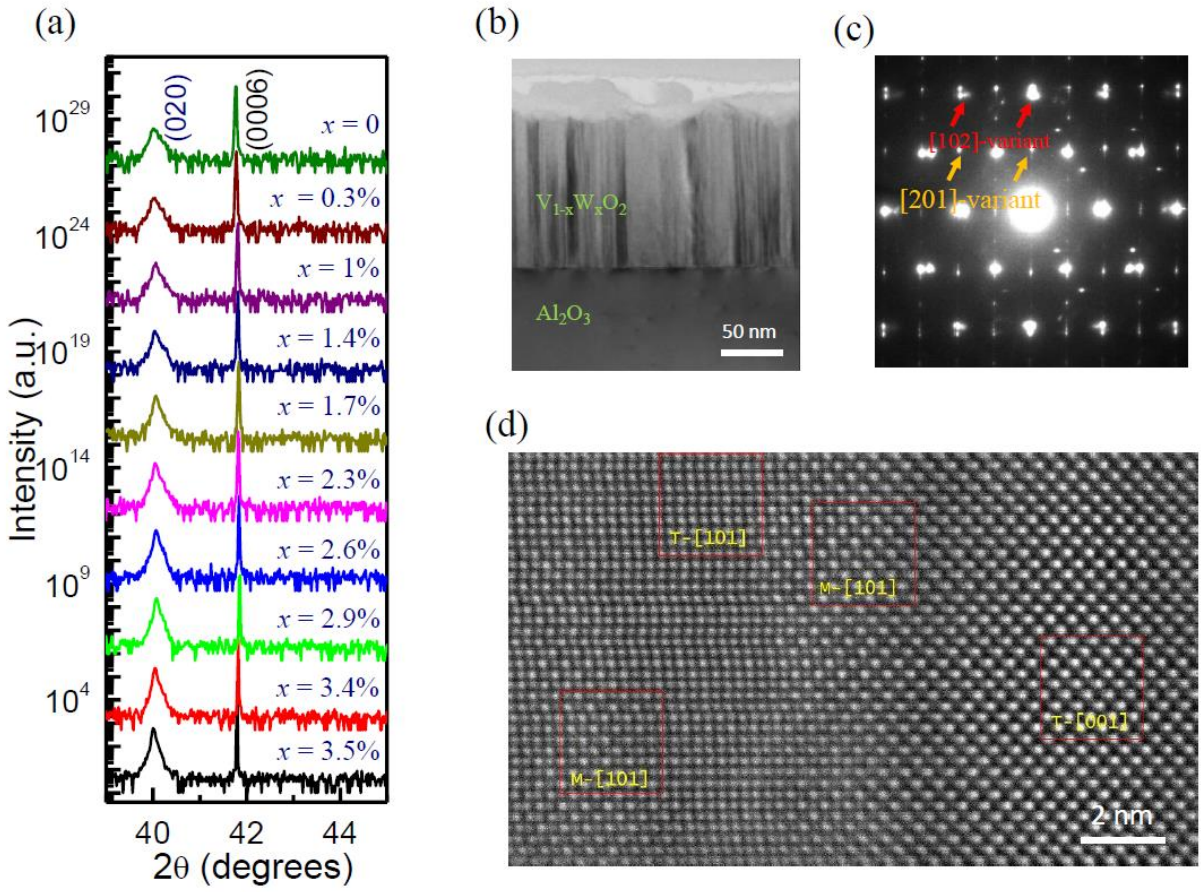

FIG. 3

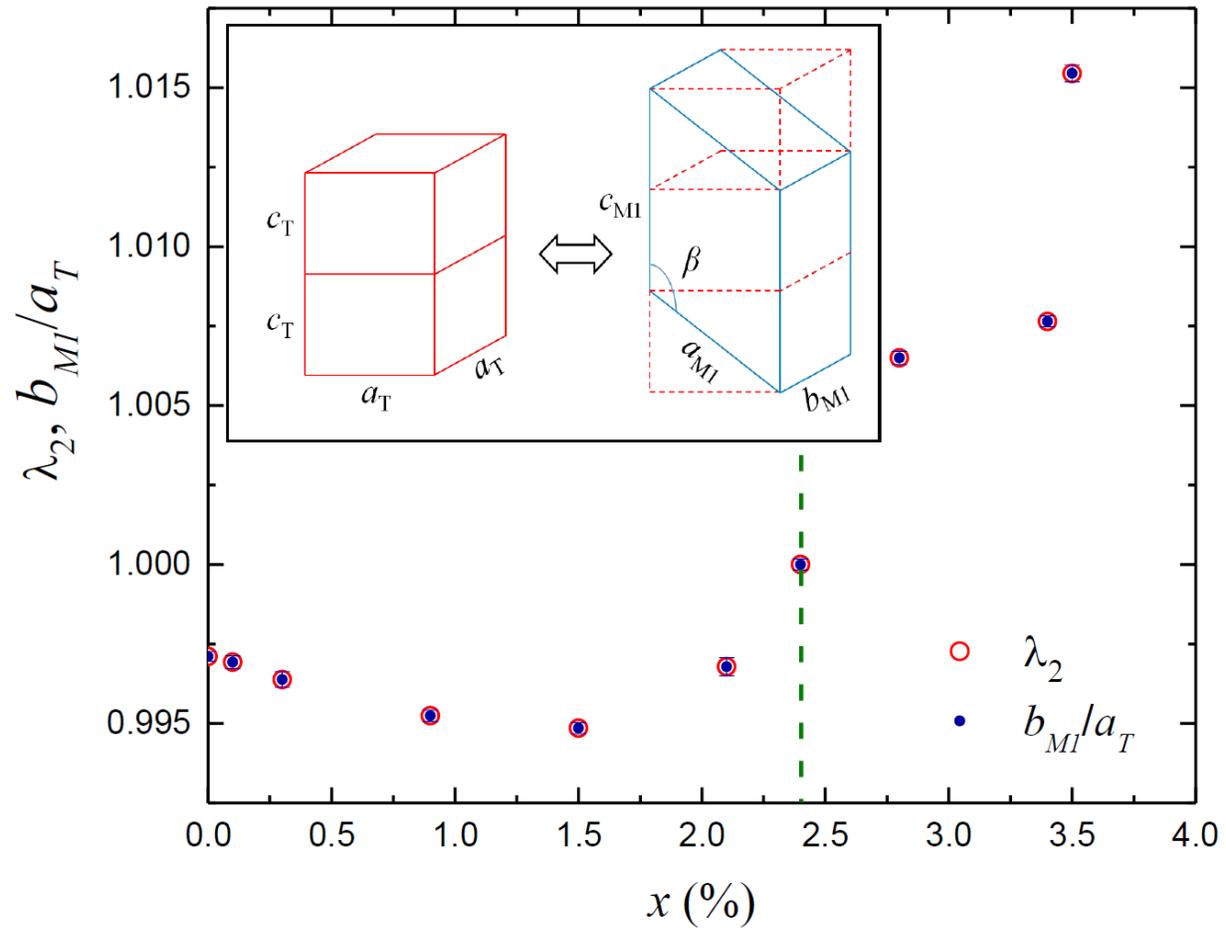

FIG. 4

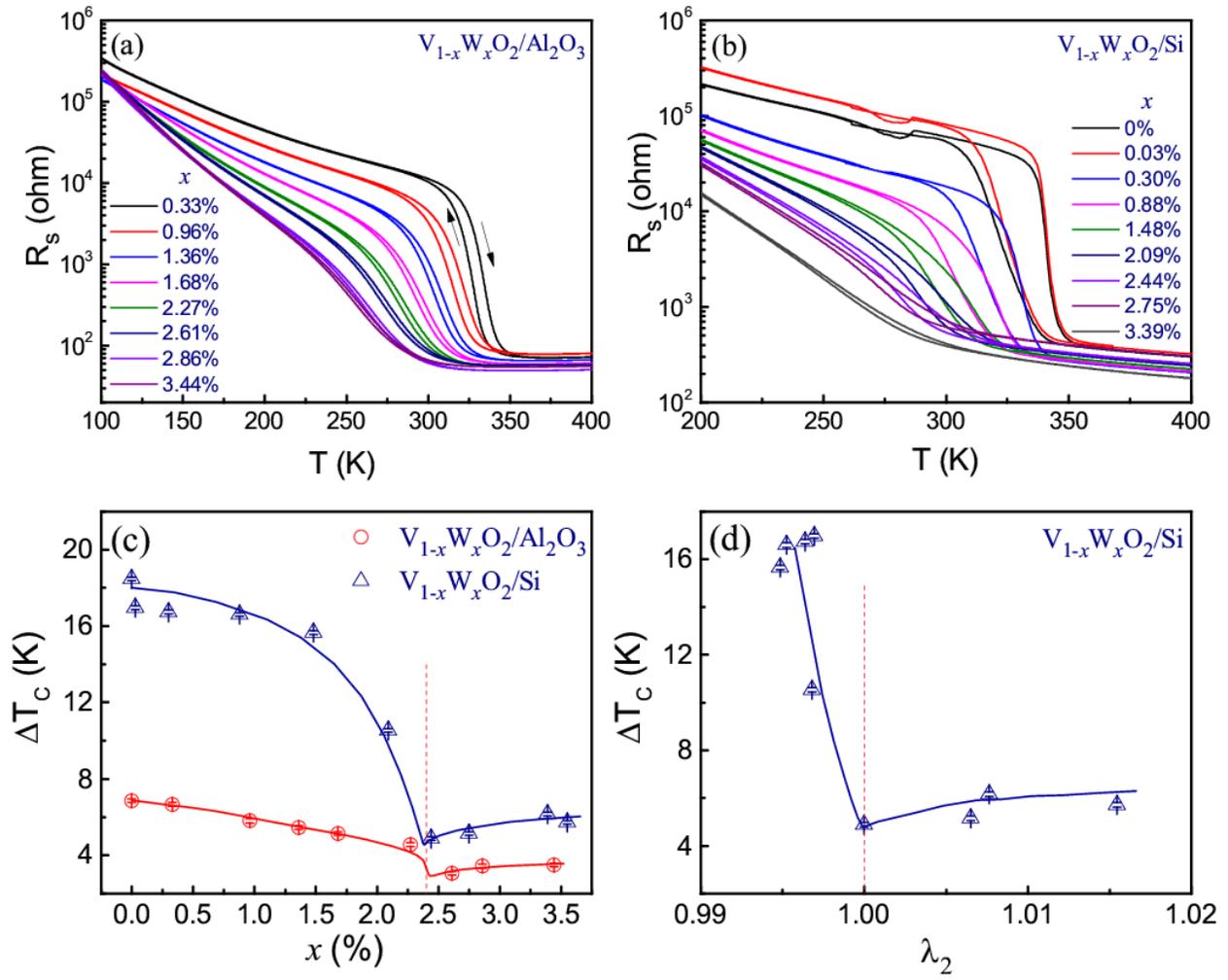

FIG. 5

| x% in $V_{1-x}W_xO_2$ | T (K) | % Phases | | Lattice constant-Monoclinic Phase | | | | | Lattice constant-Tetragonal phase | | |
|---|---|---|---|---|---|---|---|---|---|---|---|
| | | %Mono | %Tetra | a (Å) | b (Å) | c (Å) | beta (deg.) | volume (Å³) | a (Å) | c (Å) | volume (Å³) |
| 0 | 300 | 100 | 0 | 5.8751(0.0073) | 4.5217(0.0021) | 5.4861(0.0022) | 124.359(0.016) | 120.313(0.082) | | | |
| 0.1(0.1) | 300 | 100 | 0 | 5.8763(0.0015) | 4.5221(0.0021) | 5.4920(0.0022) | 124.279(0.016) | 120.591(0.083) | | | |
| 0.3(0.1) | 300 | 100 | 0 | 5.8999(0.0014) | 4.5225(0.0020) | 5.5037(0.0021) | 124.535(0.015) | 120.973(0.079) | | | |
| 0.9(0.1) | 300 | 65.6(0.15) | 34.4(0.15) | 5.9201(0.0016) | 4.5208(0.0019) | 5.5105(0.0023) | 124.634(0.019) | 121.347(0.084) | 4.5424(0.0011) | 2.86391(0.00070) | 59.091(0.033) |
| 1.5(0.1) | 300 | 62.3(0.14) | 37.7(0.14) | 5.9356(0.0016) | 4.5254(0.0020) | 5.5367(0.0023) | 125.029(0.019) | 121.783(0.087) | 4.5488(0.0012) | 2.86120(0.00070) | 59.203(0.034) |
| 2.1(0.1) | 300 | 63.50(0.1) | 36.5(0.1) | 6.0801(0.0019) | 4.5311(0.0024) | 5.5918(0.0025) | 125.944(0.023) | 124.72(0.10) | 4.5457(0.0011) | 2.85593(0.00057) | 59.014(0.031) |
| 2.4(0.1) | 300 | 62.53(0.9) | 37.47(0.9) | 6.0652(0.0014) | 4.5514(0.0017) | 5.5878(0.0018) | 125.748(0.017) | 125.193(0.073) | 4.55138(0.00090) | 2.85959(0.00053) | 59.237(0.026) |
| 2.8(0.1) | 270 | 65.90(0.79) | 34.10(0.79) | 6.1421(0.0015) | 4.5784(0.0019) | 5.6338(0.0020) | 126.873(0.017) | 126.736(0.080) | 4.54883(0.00096) | 2.85830(0.00060) | 59.144(0.028) |
| 3.4(0.1) | 270 | 64.83(0.83) | 35.17(0.83) | 6.1607(0.0014) | 4.5853(0.0018) | 5.6325(0.0019) | 126.798(0.015) | 127.409(0.076) | 4.5505(0.0010) | 2.86554(0.00065) | 59.336(0.029) |
| 3.5(0.1) | 270 | 48.78(0.97) | 51.22(0.97) | 6.2953(0.0015) | 4.6135(0.0021) | 5.7095(0.0020) | 128.311(0.018) | 130.114(0.087) | 4.54333(0.00090) | 2.86259(0.00059) | 59.089(0.026) |

**Table s1.** Phase Fractional atomic-percentage determined by Rietveld refinement. Data analysis was carried out using the TOPAS software. (Monoclinic phase: ICSD #74705, space group: $P2_1/c$; Tetragonal phase: ICSD #4110, space group: $P4_2/mnm$).

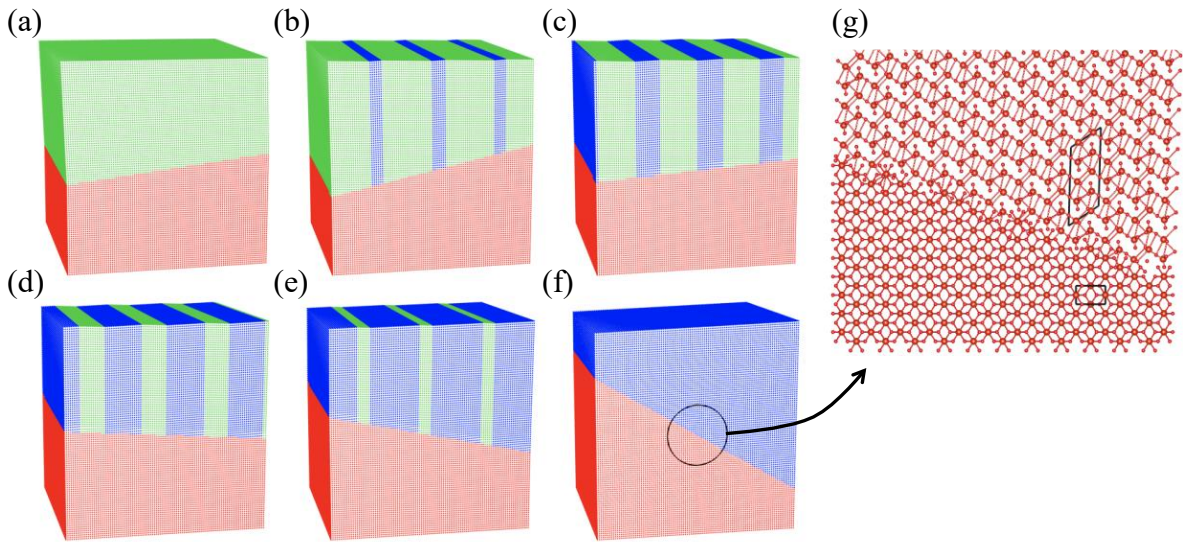

**Figure S1.** (a)-(f) Supercompatible microstructures in $V_{1-x}W_xO_2$, determined from theory evaluated at the measured lattice parameters at x = 2.4%. Front face is (100), red is tetragonal, and blue/green are two compound twinned monoclinic variants. Low energy interfaces are possible at all volume fractions of the twins and perfect unstressed interfaces are possible at volume fraction 0 or 1; (g) An atomistic view of a compatible interface (The unit cells of different variants).

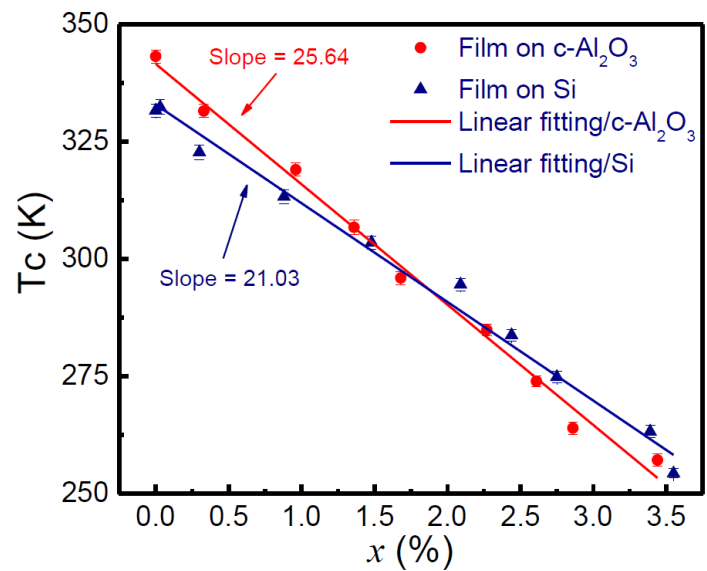

**Figure S2.** Dependence of the phase transition temperature (T$_C$) on W-concentration for samples on *c*-Al$_2$O$_3$ and Si.

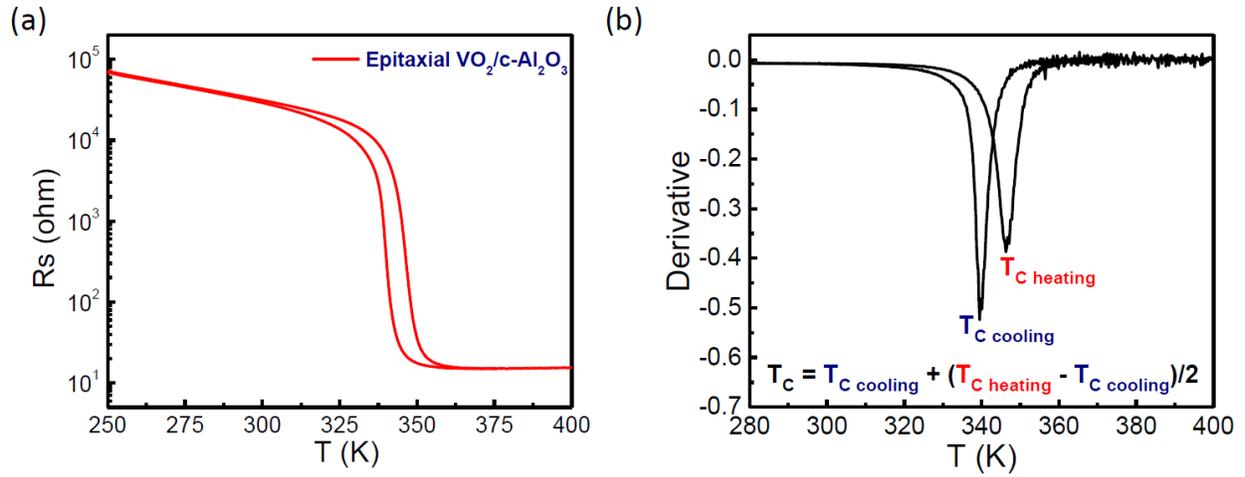

**Figure S3.** (a) The R-T curve of pure $VO_2$ on c-$Al_2O_3$. (b) The first derivative plot for log(Rs)-T from which we extract the MIT properties: transition temperature ($T_{C\ heating}$, $T_{C\ cooling}$, and $T_C$) and transition hysteresis width ($\Delta T_C = T_{C\ heating} - T_{C\ cooling}$).